\documentclass[12pt]{iopart}
\usepackage{iopams}
\usepackage{graphicx}
\input amssym.def \input amssym

\begin{document}

\title[]{Pauli graphs when the Hilbert space dimension contains a square: why the Dedekind psi function ?}

\author{Michel Planat 
%and Metod Saniga$^{2}$
}

\address{ Institut FEMTO-ST, CNRS, 32 Avenue de
l'Observatoire,\\ F-25044 Besan\c con, France. }
%\ead{michel.planat@femto-st.fr}
\vspace*{.1cm}

%\address{$^2$ Astronomical Institute, Slovak Academy of Sciences,\\ SK-05960 Tatransk\`{a} Lomnica, Slovak Republic.}
%\ead{msaniga@ta3.sk}
%\vspace*{.1cm}

\begin{abstract}
We study the commutation relations within the Pauli groups built on all decompositions of a given Hilbert space dimension $q$, containing a square, into its factors. Illustrative low dimensional examples
are the quartit ($q=4$) and two-qubit ($q=2^2$) systems, the octit ($q=8$), qubit/quartit ($q=2\times 4$) and three-qubit ($q=2^3$) systems, and so on. In the single qudit case, e.g. $q=4,8,12,\ldots$,
one defines a bijection between the $\sigma (q)$ maximal commuting sets [with $\sigma[q)$ the sum of divisors of $q$] of Pauli observables and the maximal submodules of the modular ring $\mathbb{Z}_q^2$, that arrange into the projective line $P_1(\mathbb{Z}_q)$ and a independent set of size $\sigma (q)-\psi(q)$ [with $\psi(q)$ the Dedekind psi function]. In the multiple qudit case, e.g. $q=2^2, 2^3, 3^2,\ldots$, the Pauli graphs rely on symplectic polar spaces such as the generalized quadrangles $GQ(2,2)$ (if $q=2^2$) and $GQ(3,3)$ (if $q=3^2$). More precisely, in dimension $p^n$ ($p$ a prime) of the Hilbert space, the observables of the Pauli group (modulo the center) are seen as the elements of the $2n$-dimensional vector space over the field $\mathbb{F}_p$. In this space, one makes use of the commutator to define a symplectic polar space $W_{2n-1}(p)$ of cardinality $\sigma(p^{2n-1})$, that encodes the maximal commuting sets of the Pauli group by its totally isotropic subspaces. Building blocks of $W_{2n-1}(p)$ are punctured polar spaces (i.e. a observable and all maximum cliques passing to it are removed) of size given by the Dedekind psi function $\psi(p^{2n-1})$. For multiple qudit mixtures (e.g. qubit/quartit, qubit/octit and so on), one finds multiple copies of polar spaces, ponctured polar spaces, hypercube geometries and other intricate structures.  Such structures play a role in the science of quantum information.
\end{abstract}

%\maketitle

\noindent

\section{Introduction}

The $q$-level quantum systems (also denoted $q$-dits, or qudits), and tensor products of them, possibly with a different number of levels in each factor, are basic constituents of quantum information processing. Multiple qubits, that are tensor products of two-qubit systems are routinely employed in quantum algorithms, but multiple copies of $q$-dits (with $q>2$) may turn to be more interesting in terms of self error-correction, and in relation to multipartite communication, as on the quantum Internet.  The most general system would be a mixture of multiple qudits corresponding to the factors of a integer factorization of the Hilbert space dimension as $q=\prod_{i}q_i^{p_i}$. Let us point out that, for a given dimension $q$, there exists several such factorizations, leading to distinct quantum systems. In the lowest dimensional case involving a square, one has either $q=4$ or $q=2^2$, corresponding to the single quartit and two-qubit systems, respectively. It may be convenient to use a $4$-level system (like the states of a nuclear spin $\frac{3}{2}$) to physically implement the two-qubit CNOT gate \cite{Hirayama2006}, and in some respect both systems display similar symmetries (like in the Bloch sphere representation) \cite{Planat2010}, but in general they have distinctive features (like in the Pauli group of observables and in the structure of the maximal commuting sets).

In this paper, we focus on the commutation relations of observables attached to a selected decomposition of the Hilbert space dimension $q$. The observables in a factor are defined from the action on a vector $\left|s\right\rangle$ of the $q_i$-dimensional Hilbert space of the $q_i$-dit Pauli group generated by two unitary $X$ (shift) and clock $Z$ operators via $X\left|s\right\rangle=\left|s+1\right\rangle$ and $Z\left|s\right\rangle=\omega^s \left|s\right\rangle$, with $\omega$ a primitive $q_i$-th root of unity. Then the observables in dimension $q$ are obtained by taking tensor products over the $q_i$-dimensional observable of each factor. A Pauli graph is constructed by taking the observables as vertices and a edge joining two commuting observables. Maximal sets of mutually commuting observables, i.e. maximum cliques of the Pauli graph, are used to define a point/line incidence geometry with observables as points and maximum cliques as lines. 

In recent papers, multiple qubits \cite{Planat2007,Saniga2007}, single qudits \cite{Havlicek2007,Havlicek2008,Albouy2009} and a few examples of qudit mixtures \cite{Planat2007bis} were already explored. Further work was published to clarify this earlier work dealing with symplectic polar spaces of multiple qudits \cite{Thas2009,Havlicek2009,Havlicek2010} and, in what concerns multiple qubits, its link to units in Clifford algebras \cite{Sengupta2009}, to Lie algebras \cite{Rau2009} and to a class of singular curves in phase space \cite{Klimov2010}. Prior to the advent of quantum information science, the incidence properties of the $q$-dimensional geometry and the relations to Clifford algebras were published in \cite{Shaw1989, Shaw1990}. The link of mutual unbiasedness to the general theory of angular momentum is explored in \cite{Kibler2009}, and its link to Feymann's path integral may be found in \cite{Tolar2009}.

In this paper, we focus on quantum systems of Pauli observables defined over the Hilbert space of dimension $q$ containing a square. In the single qudit case, studied in Sec. 2, the maximal mutually commuting sets of observables in the Hilbert space of dimension $q$ are mapped bijectively to the maximal submodules over the ring $\mathbb{Z}_q$ \cite{Havlicek2007,Havlicek2008}. If $q$ contains a square, there are $\psi(q)=q\prod_{p|q}(1+\frac{1}{p})$  points on the projective line $P_1(\mathbb{Z}_q)$ (in the Dedekind finction $\psi(q)$, the product is taken over all primes $p$ dividing $q$) and the remaining $\sigma(q)-\psi(q)\ne 0$ independent points (with $\sigma(q)$ the sum of divisors function) is playing the role of a reference frame and possess their own modular substructure. The number theoretical properties of the modular ring $\mathbb{Z}_q$ are used to count the cardinality of the symplectic group $\mbox{Sp}(2,\mathbb{Z}_q)$ \cite{Albouy2009,Vourdas2010,Novotny2005}. In Sec. 3, we remind the established results concerning the point/line geometries attached to multiple qudit systems in dimension $p^n$, that symplectic polar spaces $W_{2n-1}(p)$ of order $p$ and rank $n$ govern the commutation structure of the observables. Here, the number theoretical functions $\sigma(p^{2n-1})$ and $\psi(p^{2n-1})$ are found to count the number of observables in the symplectic polar space and in the {\it punctured} polar space, respectively. In Sec. 4, we study composite systems when at least one of the factors $q_i$ of the Hilbert space dimension is a square. It is shown, that the non-modularity leads to a natural splitting of the Pauli graph/geometry into several copies of basic structures such as polar spaces, punctured polar spaces and related hyperdimensional structures. 

A few properties of the structures we have checked are in table 1. Details are given in the subsequent sections.

Most calculations are performed on Magma \cite{Magma}. High dimensional computations have been made possible thanks to the supercomputer facilities of the M\'esocentre de calcul at University of Franche-Comt\'e.

\begin{table}[ht]
\begin{center}
\footnotesize
\begin{tabular}{|r|r|r|r|r|r|}
\hline
 $q$ & name [Ref.] & \# cliques & geometry & spectrum & aut. group \\
\hline

$4$ & quartit \cite{Havlicek2008,Planat2010} & $6+1$ & $P_1(\mathbb{Z}_4) ^{\dag}$ & $\{4^1,0^{3+1},-2^2\}$ &$G_{48}=\mathbb{Z}_2 \times S_4$ \\
%\hline
$2^2$& $2$-qubit \cite{Planat2007,Saniga2007} & $15$ & $GQ(2,2)$ & $\{6^1, 1^9,-3^5\}$ &$S_6 $\\
\hline
$8$ & octit \cite{Havlicek2008,Planat2010}  & $12+3$ &$P_1(\mathbb{Z}_8)$$^{\dag}$ & $\{8^1,0^{9+3},-4^2\}$ & $\mathbb{Z}_2^6 \rtimes(\mathbb{Z}_3^3 \rtimes G_{48})$\\
$2 \times 2^2 $ & qubit/quartit \cite{Planat2007bis}& $36+3$ & $3 \times GQ(2,2)'$ & $\{5^1,1^6,-1^2,-3^3\}^3$ & $G_{48}^3 \rtimes S_3$\\
$2^3$ & $3$-qubit \cite{Planat2007,Saniga2007}   & $135$ & $W_5(2)$ &$\{30^1,3^{35},-5^{27}\}$  &$\mbox{Sp}(6,2)$  \\
 \hline
$9$  & $9$-dit \cite{Havlicek2008}& $12+1$ & $P_1(\mathbb{Z}_9)$$^{\dag}$ &$\{9^1,0^{8+1 },-3^3\}$ &$G_{648} \rtimes G_{48}$\\
$3^2$ & $2$-qutrit  \cite{Planat2007}  & $40$ & $GQ(3,3)$ &$\{25^1,5^{24},-1^{40},-7^{15}\}$  & $\mathbb{Z}_2^{40}.W(E_6)$ \\
 \hline
$12$ & $12$-dit \cite{Havlicek2008} & $24+4$& $P_1(\mathbb{Z}_{12})$$^{\dag}$ &$\{12^1,2^6,0^{12+4},-4^3,-6^2\}$ &$\mathbb{Z}_2^{12} \rtimes  G_{144}$   \\
 $3\times 4$ &qutrit/quartit & $24+4$& as above& as above& as above \\
 $2^2\times 3$ & $2$-qubit/qutrit &$60$ &  $4\times GQ(2,2)$&$\{6^1,1^9,-3^5\}^4$  & $S_6^4 \rtimes S_4$\\
 \hline
 $16$& $16$-dit \cite{Havlicek2008} & $24+7$& $P_1(\mathbb{Z}_{16})$$^{\dag}$ & $\{16^1,0^{21+7},-8^2\}$ & $A_8^3 \rtimes G_{48}$ \\
 $2 \times 8$& qubit/octit &$72+15$ &$6 \times GQ(2,2)'$ &$\{5^1,1^6,-1^2,-3^3\}^6$ & $G_{48}^6 \rtimes S_6$\\
 $4 \times 4$& $2$-quartit & $120+30+1$ &$15$-cube & $\{-3^1,3^1,-1^3,1^3\}^{15}$ & $G_{48}^{15} \rtimes S_{15}$\\
 $2^2 \times 4$& $2$-qubit/quartit  & $360+15$ & $3 \times W_5(2)'$ &$\{13^1,5^{25},3^9,-1^{70},-5^5,-7^{10}\}^3$ & $(\mathbb{Z}_2^5 \rtimes S_6)^3\rtimes S_3$\\
 $2^4$& $4$-qubit \cite{Planat2007,Saniga2007} & $2295$&$W_7(2)$ &$\{126^1,7^{135},-9^{119}\}$ &$\mbox{Sp}(8,2)$ \\
 \hline
 $18$ &$18$-dit \cite{Havlicek2008} & $36+3$ &$P_1(\mathbb{Z}_{18})$$^{\dag}$ &$\{18^1,3^6,0^{24+3},-6^3,-9^2\}$ &$\mathbb{Z}_3^{12}\rtimes (\mathbb{Z}_2^{12}\rtimes G_{144}) $\\
 $2 \times 9 $ & qubit/$9$-dit &$36+3$ & as above& as above& as above\\
 $2 \times 3^2$& $2$-qutrit/qubit \cite{Planat2007bis}&$120$ & $3 \times GQ(3,3)$ & $\{12^1,2^{24},-4^{15}\}^3$& $W'(E_6)^3.G_{48}$ \\
 \hline
 $24$ & $24$-dit \cite{Havlicek2008} & $48+12$ & $P_1(\mathbb{Z}_{24})$$^{\dag}$ & $\{24^1,4^6,0^{36+12},-8^3,-12^2\}$ & $G_{2^{24}3^{12}}\rtimes(\mathbb{Z}_2^{12} \rtimes  G_{144})$\\
 $2 .3 . 4$ & qubit/qutrit/quartit & $144+12$ & see Sec. 4 & see Sec. 4 & \\ 
 $2^3 \times 3$ & $3$-qubit/qutrit & $540$ & $4 \times W_5(2) $ & $\{56^1,14^{15},2^{35},-4^{84}\}^4$ & $\mbox{Sp}(6,2)^4.S_4$ \\
 \hline
\end{tabular}
\label{structures}
\normalsize
\caption{The main properties of the studied Pauli graphs. The first and second column gives the selected decomposition of $q$ and the name of the corresponding Pauli system, respectively. Third column represents the number of  maximal sets of mutually commuting observables of size $q-1$ (i.e. the number of maximum cliques in the corresponding Pauli graph) and how it splits into two numbers of geometrical significance explained in the paper. The fourth column provides a geometry that may be identified. The fifth column provides the spectrum of the Pauli graph, that of its dual geometry or that of an important subgraph, depending on context (see the corresponding section for details). The automorphism group of the selected geometry is given in the last column. The notation $S_n$, $A_n$ and $D_n$ is for the symmetric, alternating and dihedral group, respectively. Symbols $\times$, $\rtimes$ and $.$ are for the direct, semidirect and not semidirect products of groups, respectively. 

The notation $W_{2n-1}(p)$ is for the symplectic polar space of order $p$ and rank $n$ \cite{Planat2007,Saniga2007}. The polar space $W_3(2)$ is the generalized (self-dual) quadrangle of order two GQ(2,2), also called the doily. The notation $W_{2n-1}(p)'$ means the polar space $W_{2n-1}(p)$ minus a perp-set (i.e. a point and the maximum cliques passing through it).  Whenever multiple polar spaces are featured in the table, it means that we are dealing with the mutual incidence of cliques at multiple points (see Sec. 4 for details).\\ 
$^{\dag}$ The incidence geometry is associated to the maximum cliques of the Pauli graph and the spectrum is that of all cliques (see Sec. 2 for details).}
\end{center}
\end{table}
%

%%%%%%%%%%%%%%%%%%%%%%%%%%%%%%%%%%%%%%%

\section{Pauli graph/geometry of a single qudit}

A single qudit is defined by a Weyl pair $(X,Z)$ of {\it shift} and {\it clock} cyclic operators satisfying 
\begin{equation}
ZX-\omega XZ =0,
\label{Weylpair}
\end{equation}
where $\omega=\exp \frac {2i\pi}{q}$ is a primitive $q$-th root of unity and $0$ is the null $q$-dimensional matrix. In the standard computational basis $\{\left|s\right\rangle, s \in {\mathbb{Z}_q}\}$, the explicit form of the pair is as follows
\begin{equation}
X=\left(\begin{array}{ccccc} 0 &0 &\ldots &0& 1 \\1 & 0  &\ldots & 0&0 \\. & . & \ldots &.& . \\. & . & \ldots &.& . \\0& 0 &\ldots &1 & 0\\ \end{array}\right),~~ Z= \mbox{diag}(1,\omega,\omega^2,\ldots,\omega^{q-1}).
\label{Paulis}
\end{equation}

The Weyl pair generates the single qudit Pauli group $\mathcal{P}_q=\left\langle X,Z\right\rangle$, of order $q^3$, where each element may be written in a unique way as $\omega^aX^bZ^c$, with $a,b,c \in \mathbb{Z}_q$.

It will be shown in this section that the study of commutation relations in a arbitrary single qudit system may be based on the study of symplectic modules over the modular ring $\mathbb{Z}_q^2$, and conversely that the elegant number theoretical relations underlying the {\it isotropic} lines of $\mathbb{Z}_q^2$ have their counterpart in the maximal commuting sets of a qudit system. Our results may be found in various disguises in several publications where the proofs are given \cite{Havlicek2008,Albouy2009,Vourdas2010,Novotny2005}.

Let us start with the Weyl pair property (\ref{Weylpair}) and write the group theoretical commutator as $\left[X,Z\right]=XZX^{-1}Z^{-1}=\omega^{-1} I_q$ (where $I_q$ is the $q$-dimensional identity matrix), so that one gets the expression  
\begin{equation}
\left[\omega^aX^bZ^c,\omega^{a'}X^{b'}Z^{c'}\right]=\omega^{cb'-c'b}I_q,
\label{transfer}
\end{equation}
meaning that two elements of $\mathcal{P}_q$ commute if only if the determinant $\Delta=\mbox{det}\left(\begin{array}{cc} b' &b \\c'& c \\ \end{array}\right)$ vanishes.
Two vectors such that their symplectic inner product $\left[(b',c').(b,c)\right] =\Delta=b'c-bc'$ vanishes are called perpendicular. Thus, from (\ref{transfer}), one can transfer the study of commutation relations within the group $\mathcal{P}_q$ to the study of perpendicularity of vectors in the ring $\mathbb{Z}_q^2$ \cite{Havlicek2008}.

From (\ref{transfer}), one gets the important result that the set $\mathcal{P}_q'$ of commutators (also called the derived subgroup) and the center $Z(\mathcal{P}_q)$ of the Pauli group $\mathcal{P}_q$ are identical, and one is led to the isomorphism
\begin{equation}
(\mathcal{P}_q/Z(\mathcal{P}_q),\times) \cong (\mathbb{Z}_q^2,+),
\label{iso}
\end{equation}
i.e. multiplication of observables taken in the central quotient $\mathcal{P}_q/Z(\mathcal{P}_q)$ transfers to the algebra of vectors in the $\mathbb{Z}_q$-module $\mathbb{Z}_q^2$ endowed with the symplectic inner product \lq\lq .".

\subsection*{Isotropic lines of the lattice $\mathbb{Z}_q^2$}

Let us now define a {\it isotropic line} as a set of $q$ points on the lattice $\mathbb{Z}_q^2$ such that the symplectic product of any two of them is $0 (\mbox{mod}~ q)$. From (\ref{iso}), to such an isotropic line corresponds a maximal commuting set in  $\mathcal{P}_q/Z(\mathcal{P}_q)$.

Taking the prime power decomposition of the Hilbert space dimension as $q=\prod_i p_i^{s_i}$, it is shown in (18) of \cite{Albouy2009} that the number of isotropic lines of the lattice $\mathbb{Z}_q^2$ reads 
\begin{equation}
\eta(q)=\prod_i \frac{p_i^{s_i+1}-1}{p_i-1}\equiv \sigma (q),
\label{divisor1}
\end{equation}
where $\sigma(q)$ denotes the sum of divisor function \footnote{The identification of $\eta(q)$ to $\sigma(q)$ is not provided in \cite{Albouy2009}. However, it is easy to see that the factors  in (\ref{divisor1}) are $\frac{p_i^{s+1}-1}{p_i-1}=1+p_i+p_i^2+\cdots + p_i^s=\sigma(p_i^s)$ and, since $\sigma(q)$ is multiplicative, (\ref{divisor1}) immediately follows. Similarly, the identification of $\eta(q;x)$ to $\sigma(\tilde{q}(x))$ given in (\ref{divisor2}) is easy to establish.}

It may be checked from table 1 (colum 3), that the number of maximum cliques in the Pauli graph of $\mathcal{P}_q$ [i.e. the number of maximal commuting set in $\mathcal{P}_q/Z(\mathcal{P}_q)$] in the considered single qudit decompositions $q=4,8,9,12,16$ and $18$ are $\sigma(4)=1+2+4=7$, $\sigma(8)=1+2+4+8=15$, $\sigma(9)=13$, $\sigma(12)=27$, $\sigma(16)=31$ and $\sigma(18)=39$, respectively. 

Another important quantity is the number $\eta(q;x)$ of isotropic lines through a given point $x=(b,c)$ of the lattice. Denoting by $t_i=v_{p_i}(x)$ the $p_i$-valuation \footnote{The $p$-adic valuation $v_p(x)$ of a integer number $x$ is the highest exponent $t$ suct that the power of prime $p^t$ divides $x$.} of $x$, it is shown in (36) of \cite{Albouy2009} that one obtains
\begin{equation}
\eta(q;x)=\prod_i \frac{p_i^{t_i+1}-1}{p_i-1}\equiv \sigma(\tilde{q}(x)),
\label{divisor2}
\end{equation}
where $\tilde{q}(x)=\prod_i p_i^{t_i} \le q$ is a {\it local} dimension defined at the selected point $x$.

\subsection*{The projective line $P_1(\mathbb{Z}_q)$ and the symplectic group $\mbox{Sp}(2,\mathbb{Z}_q)$}

As shown in \cite{Albouy2009}, a isotropic line of $\mathbb{Z}_q^2$ corresponds to a {\it Lagrangian submodule}, i.e. a maximal module such that the perpendicular module $M^{\perp}=M$. Let us now specialize to Lagrangian submodules that are {\it free cyclic submodules}
\begin{equation}
\mathbb{Z}_q(b,c)=\left\{(ub,uc)|u \in \mathbb{Z}_q\right\},
\label{module}
\end{equation}
for which the application $u \rightarrow (ub,uc)$ is injective. Not all Lagrangian submodules are free cyclic submodules. A point $x=(b,c)$ such that $\mathbb{Z}_q(b,c)$ is free is called an {\it admissible point}, and the set of admissible points is called the projective line
\begin{equation}
\mathbb{P}_1(\mathbb{Z}_q)=\left\{\mathbb{Z}_q(b,c)|(b,c) ~\mbox{is}~\mbox{admissible}\right\}.
\label{proj1}
\end{equation}
Following theorem 5 in \cite{Havlicek2008}, the number of points of the projective line is
\begin{equation}
|\mathbb{P}_1(\mathbb{Z}_q)|=\prod_i (p_i^{s_i}+p_i^{s_i-1})\equiv \psi(q),
\label{proj1card}
\end{equation}
where $\psi(q)=q \prod_{p|q}(1+\frac{1}{p})$ and the product is taken over all primes $p$ dividing $q$ \footnote{As for the relation (\ref{divisor1}), the identification of $|\mathbb{P}_1(\mathbb{Z}_q)|$ to the Dedekind psi function $\psi(q)$ is not provided in \cite{Havlicek2008}. The proof is easy to establish since $\psi(q)$ is a multiplicative function}. Note that one has $\psi(q) \le \sigma(q)$, where the equality holds if $q$ is square-free integer.

In the considered single qudit decompositions $q=4,8,9,12,16$ and $18$, that contain a square, one gets $\psi(4)=4(1+\frac{1}{2})=6,~ \psi(8)=8(1+\frac{1}{2})=12,~ \psi(9)=12,~\psi(12)=24,~\psi(16)=24,~\psi(18)=36$, as it it is also shown in table 1 (column 3).

Then, still using theorem 5 in \cite{Havlicek2008}, the number of points of the projective line containing a selected vector $x=(b,c)$ of the lattice reads as 

\begin{equation}
|\mathbb{P}_1(\mathbb{Z}_q;x)|=\psi(\tilde{q}(x)),
\label{proj2card}
\end{equation}
where $\tilde{q}(x)$ is the local dimension introduced in (\ref{divisor2}).

As for the projective line $\mathbb{P}_1(\mathbb{Z}_q)$, the symplectic group $\mbox{Sp}(2,\mathbb{Z}_q)$ contains interesting number theoretical features. 

We defined an admissible vector $(b,c)$ as one leading to a point of the projective line $\mathbb{P}_1(\mathbb{Z}_q)$. If $q=p^s$, there are $p^{2s}-p^{2(s-1)}$ admissible vectors and, for arbitrary dimensions $q=\prod_i p_i^{s_i}$, the number of admissible vectors is
\begin{equation}
q^2 \prod_i (1-\frac{1}{p^2})=\phi(q)\psi(q)=J_2(q),
\label{admis}
\end{equation}
where $\phi(q)=q\prod_i (1-\frac{1}{p_i})$ is the Euler totient function and $J_2(q)$ is known as the Jordan totient function.

Following the same line of reasoning than (\ref{divisor2}) and (\ref{proj2card}), one may also define a finer structure of admissibility from the number $J_2(\tilde{q}(x))$, with $\tilde{q}(x)=\prod_i p_i^{t_i} \le q$ is the local dimension. If $q=p^s$, one has $\tilde{q}(x)=q$ and the structure is simpler than in the composite case such as $q=12$ and $q=18$.

The symplectic group $\mbox{Sp}(2,\mathbb{Z}_q)$  is built from all matrices $\left(\begin{array}{cc} b' &b \\c'& c \\ \end{array}\right)$ such that $(b,c)$ is an admissible vector and the symplectic inner product, i.e. the determinant $\Delta = b'c-bc'=1$. The cardinality of such a group is is $|\mbox{Sp}(2,\mathbb{Z}_q)|=q J_2(q)$ \cite{Vourdas2010,Novotny2005}.

\subsection*{The Pauli graph of a qudit}

In the previous subsections, we investigated the bijection between sets of operators of the Pauli group $\mathcal{P}_q$ and vectors defined over the modular ring $\mathbb{Z}_q$. More precisely, from (\ref{iso}), elements of the central quotient of the Pauli group $\mathcal{P}_q/Z(\mathcal{P}_q)$ were  mapped to vectors of the lattice $\mathbb{Z}_q^2$ and, from (\ref{divisor1}) the $\sigma(q)$ isotropic lines of $\mathbb{Z}_q^2$ were mapped to its maximal commuting sets. 

One can see these bijections in a clearer way by defining the Pauli graph $\mathcal{G}_q$ of the qudit system. The Pauli graph $\mathcal{G}_q$ is constructed by taking the observables as vertices and a edge joining two commuting observables. A maximal set of mutually commuting observables corresponds to a maximum clique of $\mathcal{G}_q$, and one further defines a point/line incidence geometry with observables as points and maximum cliques as lines. One  characterizes this geometry by creating a dual graph $\mathcal{G}_q^{\star}$ such that the vertices are the cliques and a edge joins two non-intersecting cliques. The connected component of $\mathcal{G}_q^{\star}$ corresponds to the graph of the projective line $\mathbb{P}_1(\mathbb{Z}_q)$ (as  defined in previous papers \cite{Planat2007}-\cite{Havlicek2010}).

In the subsequent sections, we shall also introduce the graph $\mathcal{G}_q^{(k)}$, in which the vertices are the maximum cliques of the Pauli graph $\mathcal{G}_q$ and a edge joins two maximum cliques intersecting at $k$ points.

\subsubsection*{The quartit system}

For the four-level system, there are $4^2-1$ observables/vertices in the Pauli graph $\mathcal{G}_4$. The $\sigma(4)=7$ maximum cliques 
\begin{eqnarray}
&cl:=\{(X^2,Z^2,Z^2X^2),(X,X^2,X^3),(X^2,Z^2X,Z^2X^3),(Z,Z^2,Z^3),\nonumber \\
&(ZX,Z^2X^2,Z^3 X^3),(Z X^2,Z^2,Z^3X^2),(ZX^3,Z^2X^2,Z^3X)\}
\label{quartit1}
\end{eqnarray}
are mapped to the following isotropic lines of $\mathbb{Z}_4^2$
\begin{eqnarray}
&il:=\{\{(0,2),(2,0),(2,2)\},\{(0,1),(0,2),(0,3)\},\{(0,2),(2,1),(2,3)\},\nonumber \\
&\{(1,0),(2,0),(3,0)\},\{(1,1),(2,2),(3,3)\},\{(1,2),(2,0),(3,2)\},\nonumber \\
&\{(1,3),(2,2),(3,1)\}\}.
\label{quartit2}
\end{eqnarray}

From the latter list, one easily observes that non-admissible vectors belong to the first line $\{(0,2),(2,0),(2,2)\}$, that corresponds to the maximum clique $(X^2,Z^2,Z^2X^2)$. The remaing  vectors in $\mathbb{Z}_4^2$ generate free cyclic submodules of the form (\ref{module}).

The sequence of degrees in $\mathcal{G}_q^{\star}$ is obtained as $(1,0,0,0,6)$, meaning that the first clique given in (\ref{quartit1}) (of degree $0$) intersects all the remaing ones, and that cliques number $2$ to $7$ in (\ref{quartit1}) (of degrees $4$) form the projective line $\mathbb{P}_1(\mathbb{Z}_4)$. Indeed, one has $|\mathbb{P}_1(\mathbb{Z}_4)|=\psi(4)=6$. There are $J_2(4)=\phi(4)\psi(4)=12$ admissible points. 

The graph $\mathcal{G}_4^{\star}$ is strongly regular, with spectrum $\{4^1,0^{3+1},-2^2\}$ (in the notations of \cite{Planat2007}); the notation $0^{3+1}$ in the spectrum means that $0^3$ belongs to the projective line subgraph and there exists an extra $0$ eigenvalue in the spectrum of $\mathcal{G}_4^{\star}$. The automorphism group of $\mathbb{P}_1(\mathbb{Z}_4)$ is found to be the direct product $G_{48}=\mathbb{Z}_2 \times S_4$ (where $S_4$ is the four-letter symmetric group).

\subsubsection*{The $12$-dit system}
The main results for all qudit systems with $4 \le q \le 18$, such that $q$ contains a square, are given in table 1. We take the composite dimension $q=2^2 \times 3$ as our second illustration. There are $12^2-1=143$ observables in the Pauli graph $\mathcal{G}_{12}$. There are $\sigma(12)=28$ maximum cliques in $\mathcal{G}_{12}$, as expected. The sequence of degrees in the dual graph $\mathcal{G}_{12}^{\star}$ is found as $(4,0,\ldots,24)$, i.e. there are four cliques of degree $0$ and the remaining $\psi(12)=24$ ones have degree $12$ (as also seen from the spectrum given in Table 1).  

Owing to the composite character of the dimension, the structure of $\mathcal{G}_{12}^{\star}$ is more complex than in the quartit case, see Fig. 1 of \cite{Havlicek2008} for a picture. All four independent cliques intersect at the three vectors $(0,6),(6,0),(6,6)$, corresponding to the three observables $X^6,Z^6,X^6 Z^6$. The remaining $24$ cliques intersect at $0,1,2,3$ or $5$ points. The automorphism group of $\mathbb{P}_1(\mathbb{Z}_{12})$ is found to be $\mathbb{Z}_2^{12} \rtimes G_{144}$, with $G_{144}=A_4 \rtimes D_6$. 

Remarkably, the automorphism groups of $\mathbb{P}_1(\mathbb{Z}_{18})$ and $\mathbb{P}_1(\mathbb{Z}_{24})$ encompass that of $\mathbb{P}_1(\mathbb{Z}_{12})$, as shown in Table 1.

\section{Pauli graph/geometry for multiple qudits}

In this section, we specialize on multiple qudits $q=p^n$, when the qudit is a p-dit (with $p$ a prime number). The multiple qudit Pauli group $\mathcal{P}_q$ is generated from the $n$-fold tensor product of Pauli operators $X$ and $Z$ [defined in (\ref{Paulis}) with $\omega=\exp(\frac{2i \pi}{p})$]. One has $|\mathcal{P}_q|=p^{2n+1}$ and the derived group $\mathcal{P}_q'$ equals the center $Z(\mathcal{P}_q)$ so that $|\mathcal{P}_q'|=p$. 

Following \cite{Saniga2007,Thas2009}, the observables of $\mathcal{P}_q/Z(\mathcal{P}_q)$ are seen as the elements of the $2n$-dimensional vector space $V(2n,p)$ defined over the field $\mathbb{F}_p$, and one makes use of the commutator  
\begin{equation}
[.,.]:~ V(2n,p) \times V(2n,p) \rightarrow  \mathcal{P}_q'
\end{equation}
to induce a non-singular alternating bilinear form on $V(2n,p)$, and simultaneously a symplectic form on the projective space $PG(2n-1,p)$ over $\mathbb{F}_p$.

Doing this, the $|V(2n,q)|=p^{2n}$ observables of $\mathcal{P}_q/Z(\mathcal{P}_q)$ are mapped to the points of the symplectic polar space $W_{2n-1}(p)$ of cardinality \footnote{The proof of this statement is given in \cite{Thas2009}. The identification of $|W_{2n-1}(p)|$ to $\sigma(p^{2n-1})$ is new in this context. It is reminiscent of (\ref{divisor1}) and has still unoticed consequences about the structure of the polar space, as explained in the sequel of the paper. For $q$-level systems (single qudits), $\sigma(q)$ and $\psi(q)$ refer to the number of isotropic lines and the number of points of the projective line, respectively (as in (\ref{divisor1}) and (\ref{proj1card})). For multiple qudits, one has $q=p^{2n-1}$ and $\sigma(q)$ and $\psi(q)$ refer to the number of points of the symplectic polar space $W_{2n-1}(p)$ and of {\it punctured polar space} $W_{2n-1}(p)'$, respectively (as in (\ref{polar1}) and (\ref{polar2})).} 
\begin{equation}
|W_{2n-1}(p)|=\frac{p^{2n}-1}{p-1} \equiv \sigma(p^{2n-1}),
\label{polar1}
\end{equation}
and two elements of $[\mathcal{P}_q/Z(\mathcal{P}_q),\times]$ commute iff the corresponding points of the polar space $W_{2n-1}(p)$ are collinear.
 
A subspace of $V(2n,p)$ is called totally isotropic if the symplectic form vanishes identically on it. The polar space $W_{2n-1}(p)$ can be regarded as the space of totally isotropic subspaces of the $(2n-1)$-dimensional projective space $PG(2n-1,p)$. Such totally isotropic subspaces, also called generators $G$, have dimension $p^n-1$ and their number is 
\begin{equation}
|\Sigma(W_{2n-1}(p))|=\prod_{i=1}^n (1+p^i).
\label{gens}
\end{equation}
Let us call a spread $S$ of a vector space a set of generators partitioning its points. The size of a spread of $V(2n,p)$ is $|S|=p^{n}+1$ and one has $|V(2n,p)|-1=|S|\times|G|=(p^{n}+1)\times (p^{n}-1)=p^{2n}-1$, as expected. 

Going back to the Pauli observables, a generator $G$ corresponds to a maximal commuting set and a spread $S$ corresponds to a maximum (and complete) set of disjoint maximal commuting sets. Two generators in a spread are mutually disjoint and the corresponding maximal commuting sets are mutually unbiased \cite{Planat2007,Planat06}.

Let us define the punctured polar space $W_{2n-1}(p)'$ as the polar space $W_{2n-1}(p)$ minus a perp-set (i.e. a point $u$ and all the totally isotropic spaces passing though it) \footnote{In the graph context the symbol ' means a puncture in the graph. It is not the same symbol as in the derived subgroup $G'$ of the group $G$.}.  Then, one gets
\begin{equation}
|W_{2n-1}(p)'|=\sigma(p^{2n-1})-\sigma(p^{2n-3})=\psi(p^{2n-1}),
\label{polar2}
\end{equation}
where $\sigma(p^{2n-3})$ is the size of a perp-set and $\psi(q)$ is the Dedekind psi function.

\subsection*{The Pauli graph of a multiple qudit}

The symmetries carried by multiple qudit systems may also be studied with Pauli graphs. We define the Pauli graph $\mathcal{G}_{p^n}$ of a multiple $p^n$-dit, as we did for the single qudit case, by taking the observables as vertices and a edge joining two commuting observables. A dual graph $\mathcal{G}^{\star}_{p^n}$ is such that the vertices are the maximum cliques and a edge joins two non-interesting cliques. One denotes $\mathcal{G}'^{\star}_{p^n}$ the corresponding graph attached to the punctured polar space. Finally, one denotes $\mathcal{G}^{(k)}_{p^n}$ the graph whose vertices are the maximum cliques of the Pauli graph $\mathcal{G}_{p^n}$ and whose edges join two maximum cliques intersecting at $k$ points.

%Finally, we denote $\mathcal{G}_{p^n}^{(k)}$, the graph whose vertices are the maximum cliques and whose edges join two maximum cliques intersecting at $k$ points.

Actual calculations have been performed for two- and three-qubits, and for two- and three-qutrits. Main results are in table 2 (see details in the corresponding subsections). Denoting $c$ the ratio between the cardinalities of  $\mbox{aut}(\mathcal{G}_{p^n}^*)$ and $\mbox{aut}(\mathcal{G}_{p^n}'^*)$, one observes that $c$ identifies to the size $\sigma(p^{2n-1})$ of the polar space $W_{2n-1}(p)$, except for the case of the $3$
-qubit system where $c$ is twice the number of cliques of the Pauli graph $\mathcal{G}_{2^3}$. Thus, the space $W_{2n-1}'(p)$ may be seen as a building block of Pauli systems. One may remind that $W_{2n-1}(p)$ {\it contracts} to $W_{2n-1}'(p)$, as the size $\sigma(p^{2n-1})$ to the size $\psi(p^{2n-1})$, that the ratio of cardinalities of their automorphism groups is the number $c$, and anticipate on the structural role of $W_{2n-1}'(p)$ in qudit mixtures, shown in Table 1 and Sec. 4.
\large
\begin{table}[ht]
\begin{center}
\begin{tabular}{|r|r|r||r|r|r|}
\hline
$q=p^n$ & name & $\mbox{aut}(\mathcal{G}_{p^n})$ & $\mbox{aut}(\mathcal{G}_{p^n}^*)$        &  $\mbox{aut}(\mathcal{G}_{p^n}'^*)$ & $c=\frac{|\mbox{aut}(\mathcal{G}_{p^n}^*)|}{|\mbox{aut}(\mathcal{G}_{p^n}'^*)|}$ \\
\hline
\hline
$2$ & qubit   & $S_3$& $S_3$ & $S_2$ & $3 \equiv \sigma(2)$            \\
\hline
$2^2$ & $2$-qubit   & $S_6$& $S_6$ & $G_{48}=\mathbb{Z}_2 \times S_4$ & $15 \equiv \sigma(2^3)$            \\
\hline
$2^3$ & $3$-qubit &$Sp(6,2)$ & $O^+(8,2)$ & $\mathbb{Z}_2^6 \rtimes A_8$ & $2 \times 135 \ne 63=\sigma(2^5)$ \\
\hline
\hline
$3$ & qutrit & $\mathbb{Z}_2^{3}\rtimes G_{48}$ & $S_4$ & $S_3$ & $ 4 \equiv \sigma(3)$ \\
\hline
$3^2$ & $2$-qutrit & $\mathbb{Z}_2^{40}.W(E_6)$ &$W(E_6)$ & $G_{648}\rtimes \mathbb{Z}_2$ & $40 \equiv \sigma(3^3)$ \\
\hline
$3^3$ & $3$-qutrit & $\mathbb{Z}_2^{364}.G$ & $G$ & $(E_{243}\rtimes \mathbb{Z}_2).W(E_6)$ & $364\equiv \sigma(3^5)$ \\
\hline
\end{tabular}
\label{syms}
\caption{Comparison of the automorphism group of the dual Pauli graph $\mathcal{G}_{p^n}^*$ and that of its {\it building block} $\mathcal{G}_{p^n}'^*$, defined  by removing a perp-set in the symplectic polar space. The ratio of sizes of both groups turns out to be the number of observables $\sigma(p^{2n-1})$ of the space, except for the case of the $3$-qubit system where it is twice the number $135$ of cliques in the Pauli graph $\mathcal{G}_{2^3}$.} 
\end{center}
\end{table}
\normalsize

\subsection*{The two-qubit system}

As already emphasized in \cite{Planat2007,Saniga2007}, the two-qubit system \lq\lq is" the symplectic polar space $W_3(2)$ [i.e. $p=n=2$ in (\ref{polar1})], alias the generalized quadrangle $GQ(2,2)$, also called {\it doily}, with $15$ points and, dually, $15$ lines (see Fig. 6 in \cite{Planat2007}). One denotes the corresponding Pauli graph as $\mathcal{G}_{2^2}$. The maximum cliques are as follows
\begin{eqnarray}
& cl:=\{ 
  \{ IX,XI,XX \},\{ IX,YI,YX \},\{ IX,ZI,ZX \},  \nonumber \\
& \{ IY,XI,XY \},\{ IY,YI,YY \},\{ IY,ZI,ZY \}, \nonumber \\ 
& \{ IZ,XI,XZ \},\{ IZ,YI,YZ \},\{ IZ,ZI,ZZ \}, \nonumber \\
& \{ XY,YX,ZZ \},\{ XY,YZ,ZX \},\{ XZ,YX,ZY \}, \nonumber \\
& \{ XZ,YY,ZX \},\{ XX,YY,ZZ \},\{ XX,YZ,ZY \}\},  
\label{clGQ22}
\end{eqnarray}
where a notation such as $IX$ means the tensor product of $I$ and $X$.

The spectrum of the (strongly regular) Pauli graph $\mathcal{G}_{2^2}$ is $\{6^2,1^9,-3^5\}$ and the automorphism group is the symmetric group $\mbox{Sp}(4,2)=S_6$.

Following definition (\ref{polar2}), ones defines the punctured polar space $W_3(2)'\equiv GQ(2,2)'$ by removing a perp-set in $GQ(2,2)$, i.e. a point as well as the totally isotropic subspaces/maximum cliques passing through it
[for the selected point $u\equiv IX$, the removed cliques are numbered 1 to 3 in (\ref{clGQ22})]. The punctured Pauli graph $\mathcal{G}_{2^2}'^*$ is as follows

\begin{eqnarray}
&GQ(2,2)'\Rightarrow \mathcal{G}'^*_{2^2}:\nonumber \\
&\mbox{spec}:= \{6^1,2^3,0^2,-2^6\},~~\mbox{aut}(\mathcal{G}'^*_{2^2}):=G_{48}=\mathbb{Z}_2 \times S_4.
\label{dGQ22}
\end{eqnarray}
The automorphism group of the graph $\mathcal{G}'^*_{2^2}$ is similar to the automorphism group obtained from the graph of the projective line $\mathbb{P}_1(\mathbb{Z}_4)$, associated to the quartit system, although the spectrum and the commutation structure are indeed not the same. In a next paper, it will be shown that both graphs are topologically equivalent to the hollow sphere.

It is already mentioned in Sec. 3 of \cite{Planat2007} that the Pauli graph $\mathcal{G}_{2^2}$ can be regarded as $\hat{L}(K_6)$ (it is isomorphic to the line graph of the complete graph $K_6$ with six vertices). Similarly, defining $K_{222}$ as the complete tripartite graph (alias the $3$-cocktail party graph, or octahedral graph), one gets $\mathcal{G}'_{2^2}=\hat{L}(K_{222})$.   

\subsection*{The two-qutrit system}

First results concerning the commutation structure of the two-qutrit system are in Sec. 5 of \cite{Planat2007} and in example 5 of \cite{Havlicek2009}. The $80$ observables of the central quotient $\mathcal{P}_q/Z(\mathcal{P}_q)$ (with $q=3^2$) are mapped to the elements of the vector space $V(4,3)$ and the commutation structure is that of the polar space $W_5(3)\equiv GQ(3,3)$ with $\sigma(3^3)=1+3+3^2+3^3=40$ elements [see (\ref{polar1})]. According to (\ref{gens}), this number coincides with the number of generators $(1+3)(1+3^2)$. The spectrum of the (regular) Pauli graph $\mathcal{G}_{3^2}$ is $\{25^1,5^{24},-1^{40},-7^{15}\}$ and the automorphism group is isomorphic to $\mathbb{Z}_2^{40}.W(E_6)$, where $W(E_6)$ is the Weyl group of the Lie algebra $E_6$. Two maximum cliques intersect at $0$ or $2$ points. The dual graph 
$\mathcal{G}_{3^2}^{\star}$ has spectrum $\{27^1,3^{15},-34^{24}\}$ and its automorphism group is $W(E_6)$. Note that $\mbox{Sp}(4,3)\cong \mathbb{Z}_2. W'(E_6)$. 

Using (\ref{polar2}), one defines the punctured polar space $W_5(3)'$, with $|W_5(3)'|=\psi(3^3)=36$ and the corresponding  graph $\mathcal{G}'^*_{3^2}$ 
\begin{eqnarray}
&GQ(3,3)'\Rightarrow \mathcal{G}'^*_{3^2}:\nonumber \\
&\mbox{spec}:= \{24^1,3^{12},0^3,-3^{20}\},~~\mbox{aut}(\mathcal{G}'^*_{3^2}):=G_{648}\rtimes \mathbb{Z}_2,
\label{dGQ33}
\end{eqnarray}
where $G_{648}\cong \mathcal{U}_{25}$ is isomorphic to a complex reflection group (number 25 in the Shephard-Todd sequence). Since $|W(E_6)|=51840=40 \times 1296$, then $GQ(3,3)$ may also be seen as $40$ copies of the building block $GQ(3,3)'$.
See Sec. VI of \cite{Briand2004} and Sec. 4.1 of \cite{Planat2009} for other occurences of the group $G_{648}$ in relation to the geometry of the $27$ lines on a smooth cubic surface.   
 
\subsection*{The three-qubit system}

For three qubits, the structure of commutation relations is that of the polar space $W_5(2)$ with $\sigma(2^5)=63$ elements and $(1+2)(1+2^2)(1+2^3)=135$ generators. The (regular) Pauli graph $\mathcal{G}_{2^3}$ has spectrum $\{30^1,3^{35},-5^{27}\}$ and $\mbox{aut}(\mathcal{G}_{2^3})=\mbox{Sp}(6,2)\cong W'(E_7)$, of order $1451520$. Two maximum cliques intersect at $0$, $1$ or $3$ points. The dual Pauli graph $\mathcal{G}_{2^3}^{\star}$ has spectrum $\{64^1,4^{84},-8^{50}\}$ and $\mbox{aut}(\mathcal{G}_{2^3}^{\star})= O^+(8,2)$. Note that $O^+(8,2)$ is related to the Weyl group of $E_8$ by the isomorphism $W(E_8)\cong \mathbb{Z}_2 . O^+(8,2)$. 

As shown in table 2, the three-qubit system is very peculiar among multiple qubit systems, having $O^+(8,2)$ as the automorphism group attached to the maximum cliques, instead of the symplectic group $\mbox{Sp}(6,2)$. 

One defines the punctured polar space $W_5(2)'$ with $|W_5(2)'|=\psi(2^5)=48$ points and the corresponding graph $\mathcal{G}'^*_{2^3}$ as follows
\begin{eqnarray}
&W_5(2)'\Rightarrow \mathcal{G}'^*_{2^3}:\nonumber \\
&\mbox{spec}:= \{56^1,4^{70},-4^{14},-8^{35}\},~~\mbox{aut}(\mathcal{G}'^*_{2^3}):=\mathbb{Z}_2^6 \rtimes A_8,
\label{dW52}
\end{eqnarray}
with $A_8$ the eight letter alternating group.

The corresponding $1$-point intersection graph of the maximum cliques has spectrum $\{56^1,14^{15},2^{35},-4^{84}\}$. As shown in Sec. 4, it occurs in the study of the $3$-qubit/qutrit system. 

Thus, the {\it number of pieces} within the automorphism group of the dual Pauli graph  $\mathcal{G}^{\star}_{2^3}$ is twice the number $135$ of maximum cliques, instead of the cardinality $\sigma(p^{2n-1})$ of the symplectic polar space $W_{2n-1}(p)$, that is given in (\ref{polar1}) \footnote
{How to explain this anomaly? Symplectic polar spaces $W_{2n-1}(p)$ are not the only type among finite polar spaces of order $p$ and rank $n$, but there are others \cite{Cameron}. Finite classical polar spaces may be of the symplectic, unitary or orthogonal type, according to the type of the reflexive sesquilinear form carried by the vector space $V(2n,p)$, alternating bilinear, Hermitian and quadratic, respectively. There are in fact six families according to their germ, viz., one symplectic, two unitary, and three orthogonal. The {\it hyperbolic} orthogonal polar space is a hyperbolic quadric $Q_{2n-1}^+(p)$ with automorphism group $O^+(2n,p)$. For such a family, the number of points is 
\begin{equation}
|Q_{2n-1}^+(p)|=\frac{(p^n-1)(p^{n-1}+1)}{p-1}
\label{hyperb1}
\end{equation}
and the number of totally isotropic subspaces is 
\begin{equation}
|\Sigma(Q_{2n-1}^+(p))|=\prod_{i=1}^n (1+p^{i-1}).
\label{hyperb2}
\end{equation}
If $n=4$, one has $|Q_7^+(p)|=\frac{(p^4-1)(p^3+1)}{p-1}=(p+1)(p^2+1)(p^3+1)=|\Sigma(W_7(p))|$, i.e. the number of points of the hyperbolic quadric $Q_7^+(p)$ coincides with the number of totally isotropic subspaces in the symplectic polar space $W_5(p)$, associated to $3$-qudit systems. But, to our great surprise, this anomaly only affects the three-qubit system.
} .

Later, in the study of the $2$-qubit/quartit system, we need the graph $\mathcal{G}^{'(3)}_{2^3}$ attached to the $3$-point intersection of the maximum cliques. The spectrum of this graph is $\{13^1,5^{25},3^9,-1^{70},-5^5,-7^{10}\}$.

To conclude this subsection, let us mention that the Weyl group $W(E_6)$ arises as the symmetry group of a subgeometry of the polar space $W_5(2)$, namely in the generalized quadrangle $GQ(2,4)$ \cite{Levay2009}. Taking the $27$ three-qubit observables shown in Fig. 3 of \cite{Levay2009}, one attaches to such a geometry a Pauli graph, that we denote $\mathcal{G}_{27}$. One gets $45$ maximum cliques of size $3$, the spectrum is $\{10^1,1^{20},-5^6\}$ and $\mbox{aut}(\mathcal{G}_{27})\cong W(E_6)$. By removing a perp-set from $GQ(2,4)$, one gets the punctured generalized quadrangle $GQ(2,4)'$. The corresponding dual Pauli graph $\mathcal{G}_{27}'^*$ has spectrum $\{32^1,2^{24},-4^{20}\}$ and automorphism group $\mathbb{Z}_2 \wr A_5$ (where $\wr$ means the wreath product of groups).

The one-point intersection graph $\mathcal{G}_{45}$ of the $45$ maximum cliques is that of the generalized quadrangle $GQ(4,2)$, the dual geometry of $GQ(2,4)$. The spectrum of $\mathcal{G}_{45}$ is $\{12^1,3^{20},-3^{24}\}$ and $\mbox{aut}(\mathcal{G}_{45})\cong W(E_6)$. The automorphism group of the punctured Pauli graph $\mathcal{G}_{45}'^*$ is isomorphic to the Weyl group $W(F_4)$ of the $24$-cell. 

 This view fits the one proposed in our paper \cite{Planat2009}.

\subsection*{The three-qutrit system}

For three qutrits [$p=n=3$ in (\ref{polar1})], the structure of the commutation relations is that of the polar space $W_5(3)$ with $\sigma(3^5)=364$ points, and [according to (\ref{polar2})] there are $(1+3)(1+3^2)(1+3^3)=1120$ generators/maximum cliques, as checked from the Pauli graph $\mathcal{G}_3^3$. Its spectrum is found to be $\{241^1,17^{195},-1^{364},-19^{168}\}$ 
 and $\mbox{aut}(\mathcal{G}_{3^3})=\mathbb{Z}_2^{364}. G$, where $|G|=|Sp(6,3)|$. Two maximum cliques of $\mathcal{G}_{3^3}$ intersect at $0$, $2$ or $8$ points. 
 
The punctured polar space $W_5(3)'$, of cardinality $|W_5(3)'|=\psi(3^5)=324$ is such that 
\begin{eqnarray}
&W_5(3)'\Rightarrow \mathcal{G}'^*_{3^3}:\nonumber \\
&\mbox{spec}:= \{702^1, 9^{780},-18^{39},-27^{260}\},~~\mbox{aut}(\mathcal{G}'^*_{3^3}):=(E_{243}\rtimes\mathbb{Z}_2).W(E_6),
\label{dW73}
\end{eqnarray}
so that $W_5(3)$ consists of $|\mbox{aut}(\mathcal{G}^*_{3^3})|/|\mbox{aut}(\mathcal{G}'^*_{3^3})|=\sigma(3^5)=364$ copies of its building block $W_5(3)'$. In (\ref{dW73}), $E_{243}$ is the extraspecial $3$-group of order $243$ and exponent $3$.

\section{Pauli graph/geometry of multiple qudit mixtures}

As before, $\mathcal{G}_q$ is the Pauli graph whose vertices are the observables and whose edges join two commuting observables. A dual graph of the Pauli graph is $\mathcal{G}^*_q$ whose vertices are the maximum cliques and whose edges join two non-intersecting cliques. In this section, we also introduces $\mathcal{G}_q^{(k)}$, the graph whose vertices are the maximum cliques and whose edges join two maximum cliques intersecting at $k$ points. 

First of all, as shown in Sec. 6 of \cite{Havlicek2007}, a qudit mixture in composite dimension $q=p_1\times p_2\times \cdots \times p_r$ ($p_i$ a prime number), identifies to a single $q$-dit. Since the ring $\mathbb{Z}_q$ is isomorphic to the direct product $\mathbb{Z}_{p_1} \times \mathbb{Z}_{p_2} \times \cdots \mathbb{Z}_{p_r}$ the commutation relations arrange as the $\sigma(q)\equiv \psi(q)$ isotropic lines of the lattice $\mathbb{Z}_q^2$, that reproduce the projective line $\mathbb{P}_1(\mathbb{Z}_q)=\mathbb{P}_1(\mathbb{Z}_{p_1}) \times \mathbb{P}_1(\mathbb{Z}_{p_2})\times \cdots \times\mathbb{P}_1(\mathbb{Z}_{p_r})$.

\subsection*{The sextit system}

The simplest non-trivial case is in dimension $q=6=2 \times 3$. The projective line may be pictured by the dual Pauli graph $\mathcal{G}_6^{\star}$ of spectrum $\{6^1,1^6,-2^3,-3^2\}$. It represents the complement of a $3 \times 4$ grid, or in graph theoretical language the complement $\hat{L}(K_{3,4})$ of the line graph over the complete bipartite graph $K_{3,4}$ (see Fig. 1 of \cite{Planat2008}). One finds $24$ maximum cliques of size $3$ in $\mathcal{G}_6^{\star}$ corresponding to the same number of non-complete sets of mutually unbiased bases. The symmetry of this new configuration is the semi-direct product $G_{144}=A_4 \rtimes D_6$ of two groups of order twelve, namely the four-letter alternating group $A_4$ and the dihedral group $D_6$. Until now, it is not known whether sets of mutually unbiased bases of size larger than three can be built \cite{Planat06,Weigert09}.

In the sequel of this section, we are interested in mixtures where at least one factor in the prime number decomposition of $q$ contains a square. A summary of the main results is in Table 3 below. 

\begin{table}[ht]
\begin{center}
\small
\begin{tabular}{|r|r|r|r|r|r|}
\hline
 $q$ & name & \# cliques & geometry &  aut. group \\
\hline
$2^2$& $2$-qubit  & $15$ & $GQ(2,2)$ & $S_6 $\\
\hline
$2 \times 2^2 $ & qubit/quartit & $36+3$ & $3 \times GQ(2,2)'$ & $G_{48}^3 \rtimes S_3$\\
$2^3$ & $3$-qubit    & $135$ & $W_5(2)$  &$\mbox{Sp}(6,2)$  \\
 \hline
$3^2$ & $2$-qutrit  & $40$ & $GQ(3,3)$   & $\mathbb{Z}_2^{40}.W(E_6)$ \\
 \hline
 $3\times 4$ &qutrit/quartit & $24+4$& as a $12$-dit& as a $12$-dit \\
 $2^2\times 3$ & $2$-qubit/qutrit &$60$ &  $4\times GQ(2,2)$& $S_6^4 \rtimes S_4$\\
 \hline
 $2 \times 8$& qubit/octit &$72+15$ &$6 \times GQ(2,2)'$  & $G_{48}^6 \rtimes S_6$\\
  $4 \times 4$& $2$-quartit & $120+30+1$ &$15$-cube &  $G_{48}^{15} \rtimes S_{15}$\\
 $2^2 \times 4$& $2$-qubit/quartit  & $360+15$ & $3 \times W_5(2)'$ & $(\mathbb{Z}_2^5 \rtimes S_6)^3\rtimes S_3$\\
 $2^4$& $4$-qubit & $2295$&$W_7(2)$  &$\mbox{Sp}(8,2)$ \\
 \hline
  $2 \times 9 $ & qubit/$9$-dit &$36+3$ & as a $18$-dit& as a $18$-dit \\
 $2 \times 3^2$& $2$-qutrit/qubit&$120$ & $3 \times GQ(3,3)$ & $W'(E_6)^3.G_{48}$ \\
 \hline
 $2 \times 3 \times 4$ & qubit/qutrit/quartit & $144+12$ & see Sec. 4 & see Sec. 4  \\ 
 $2^3 \times 3$ & $3$-qubit/qutrit & $540$ & $4 \times W_5(2)$  & $\mbox{Sp}(6,2)^4.S_4$ \\
 \hline
\end{tabular}
\label{compositess}
\normalsize
\caption{Geometry of qudit mixtures: an excerpt from table 1 which emphasizes the role played by symplectic polar spaces $W_{2n-1}(p)$ and their punctured part $W'_{2n-1}(p)$.  }
\end{center}
\end{table}

\subsection*{The two-qubit/qutrit system}

The Pauli graph of the two-qubit/qutrit system contains $143$ vertices and $60$ maximum cliques. The incidence graph of the maximum cliques is found to reproduce the projective line over the ring $\mathbb{F}_4\times \mathbb{Z}_2 \times \mathbb{Z}_3$ \cite{Planat2007bis} and the spectrum of the dual Pauli graph $\mathcal{G}_{2^2 \times 3}^{\star}$ is $\{24^1,6^5,2^{27},-2^{15},-6^9,-8^3\}$. Maximum cliques of the Pauli graph intersect each other at $0$, $1$, $2$ or $5$ points. In $\mathcal{G}_{2^2 \times 3}^{\star}$, there are $480$ maximum cliques of size $3$ and $720$ maximum cliques of size $4$, to which one can attach the same number of non-complete sets of mutually unbiased bases. 

An interesting subgeometry of the two-qubit/qutrit system is found by taking the incidence graph $\mathcal{G}_{2^2 \times 3}^{(5)}$ of maximum cliques of the Pauli graph intersecting each other at $5$ points. The spectrum of this graph is $\{6^1,1^9,-3^5\}^4$ corresponding to four {\it copies} of the doily $GQ(2,2)$ [alias $\hat{L}(K_6)$]. The automorphism group of this geometry is $S_6^4 \rtimes S_4$. Similarly, the spectrum of the incidence graph for maximum cliques intersecting at two points is $\{8^1,2^{5},-2^{9}\}^4$, that represents four copies of the triangular graph $L(K_6)$. Thus, the doily is a {\it basic constituent} of the two-qubit/qutrit system and builds up its commutation structure, as one may have expected.  

\subsection*{The two-qutrit/qubit system}

The Pauli graph of the two-qutrit/qubit system contains $323$ vertices and $120$ maximum cliques. The incidence graph of the maximum cliques reproduces the projective line over the ring $\mathbb{F}_9\times \mathbb{Z}_2 \times \mathbb{Z}_3$ and the spectrum of the dual Pauli graph $\mathcal{G}_{2 \times 3^2}^{\star}$ is $\{54^1,6^{15},3^{48},-3^{30},-6^{24},-27^2\}$. It contains $19440$ cliques of size three. Maximum cliques of the Pauli graph intersect each other at $0$, $1$, $2$, $5$ or $8$ points.

An interesting subgeometry of the two-qutrit/qubit system is found by taking the incidence graph $\mathcal{G}_{2 \times 3^2}^{(5)}$ of maximum cliques of the Pauli graph intersecting each other at $5$ points. The spectrum of this graph is $\{12^1,2^{24},-4^{25}\}^3$ corresponding to three {\it copies} of the dual graph $\mathcal{G}_{3^2}^{\star}$ of the generalized quadrangle $GQ(3,3)$.

\subsection*{The qubit/quartit and qubit/octit systems}

Let us start with the lowest case of a mixture where a factor is not a prime: the qubit/quartit system living in the Hilbert space dimension $q=2 \times 4$. As shown in Table 1, the maximum cliques studied from the dual Pauli graph $\mathcal{G}^{\star}_{2 \times 4}$ split into two parts, that are a set of $3$ independent (non-intersecting) cliques and a connected component of $36$ cliques. The single qudit in dimension $q=18$ has a similar splitting since $\sigma(18)=39$ and $\psi(18)=36$.

Let us denote $\mathcal{G}^{\star(c)}_{2 \times 4}$ the connected subgraph of the dual Pauli graph. Its spectrum is again that $\{16^1,0^{27},-8^4,4^4\}$ of a regular graph. Maximum cliques  of the Pauli graph intersect each other at $0$, $1$ or $3$ points and there is a subgeometry of the qubit/quartit system found by taking the incidence graph of maximum cliques interesting at $3$ points. The spectrum of this latter graph $\mathcal{G}_{2 \times 4}^{(7)}$ is $\{5^1,1^6,-1^2,-3^3\}^3$ corresponding to three copies of the punctured Pauli graph associated to $GQ(2,2)'$ [see (\ref{dGQ22})\footnote{The spectrum $\{5^1,1^6,-1^2,-3^3\}$ is that of the complement of the graph $\mathcal{G}_{2^2}'^{\star}$, displayed in (\ref{dGQ22}).}]. The automorphism group of this $3$-point incidence graph is $G_{48}^3 \rtimes S_3$, where $G_{48}$ is the automorphism group of $GQ(2,2)'$. 

Similarly, one considers the qubit/octit system living in the Hilbert space dimension $q=2 \times 8$. As shown in Table 1, the maximum cliques studied from the Pauli graph $\mathcal{G}^{\star}_{2 \times 8}$ split into two parts, that are a set of $15$ independent (non-intersecting) cliques and a connected component of $72$ cliques. Here, there exists no single qudit with such a splitting. Let us denote $\mathcal{G}^{\star(c)}_{2 \times 8}$ the connected subgraph of the dual Pauli graph. Its spectrum is $\{32^1,8^4,0^{63},-16^4\}$. Maximum cliques of the Pauli graph intersect each other at $0$, $1$, $3$ or $7$ points and there is a subgeometry of the qubit/octit system found by taking the incidence graph $\mathcal{G}_{2 \times 8}^{(7)}$ of maximum cliques interesting at $7$ points. 
The spectrum of this latter graph corresponds to six copies of of $GQ(2,2)'$. The automorphism group of this latter configuration is found to be the semidirect product of groups $G_{48}^6 \rtimes S_6$.

\subsection*{The two-qubit/quartit system}

The two-qubit/quartit system corresponds to the decomposition $q=2^2 \times 4$ of the Hilbert space dimension. We find $375$ maximum cliques in the Pauli graph, that split as $360+15$. The $360$ maximum cliques of the Pauli graph arrange each other in $7$-tuples. The graph spectrum of this latter graph $\mathcal{G}_{2^2 \times 4}^{(7)}$ is  $\{13^1,5^{25},3^9,-1^{70},-5^5,-7^{10}\}^3$. Thus, this is the geometry is related to three copies of the punctured polar space $W_5(2)'$.

\subsection*{The two-quartit  system}

The two-quartit system corresponds to the decomposition $q=4 \times 4$ of the Hilbert space dimension. The Pauli graph $\mathcal{G}_{4 \times 4}$ contains $151$ maximum cliques. The connected subgraph $\mathcal{G}_{4 \times 4}^{\star (c)}$ of the dual graph $\mathcal{G}_{4 \times 4}^{\star}$ corresponds to $120$ maximum cliques of the Pauli graph that intersect each other at $0$, $1$, $3$ or $7$ points. The graph $\mathcal{G}_{4 \times 4}^{(7)}$ featuring the intersection of the $120$ maximum cliques at $7$ points has spectrum $\{-3^1,3^1,-1^3,1^3\}^{15}$, that corresponds to $15$ copies of the cube graph. The automorphism group $G_{48}$ of the cube graph is similar to that of the punctured generalized quadrangle $GQ(2,2)'$. The automorphism group of the selected geometry $\mathcal{G}_{4 \times 4}^{(7)}$ is found to be $G_{48}^{15} \rtimes S_{15}$.

The remaining $31$ cliques intersect each other at $3$ or $7$ points. The $3$-clique intersection graph still splits into a isolated clique and a connected component of $30$ maximum cliques. The connected component, of spectrum $\{28^1,0^{15},-2^{14}\}$, is the $15$-cocktail party graph, i.e. the dual graph of the $15$-hypercube graph.

\subsection*{Pauli systems in dimension $24$}

A few results for dimension $24$ are collected at the bottom of table 1. Here, a transition towards a more complex behavior occurs. In smaller dimensional cases, all maximum cliques of the Pauli graph $\mathcal{G}_q$ have size $q-1$, while in dimension $24$, in all the three cases explored ($24$-dit, qubit/qutrit/quartit and $3$-qubit/qutrit), they may also have dimension $24-1$ and $24+1$. 

The $24$-dit system follows the rules established in Sec. 2, as expected. The automorphism group of the projective line $\mathbb{P}_1(\mathbb{Z}_{24})$ is found to be $G_{2^{24}3^{12}}\rtimes(\mathbb{Z}_2^{12} \rtimes  G_{144})$, where $G_{2^{24}3^{12}}$ is the product of two non elementary abelian groups of order $2^{24}$ and $3^{12}$.

The maximum cliques of the qubit/qutrit/quartit Pauli graph $\mathcal{G}_{2\times 3 \times 4}$ split as $144+12$ (as for a single $q$-dit with $q=99$, since $\sigma (99)=156$ and $\psi(99)=144$). The maximum cliques of the Pauli graph $\mathcal{G}_{2\times 3 \times 4}$  intersect each other at $0$, $1$, $2$, $3$, $5$, $7$ and $11$ points. The graph  $\mathcal{G}_{2\times 3 \times 4}^{(7)}$ of intersection at $7$-tuples has the spectrum $\{3,-1^3\}^{36}$. Similarly, one gets $\mbox{spec}(\mathcal{G}_{2\times 3 \times 4}^{(3)})=\{15^1,3^{15},1^6,-1^{18},-3^2,-9^3,-5^3\}^3$ and $\mbox{spec}(\mathcal{G}_{2\times 3 \times 4}^{(2)})=\{16^1,0^{27},-8^4,4^4\}^3$.

The Pauli graph $\mathcal{G}_{2^3 \times 3}$ of the $3$-qubit/qutrit system contains $540$ maximum cliques. There are $|\mbox{Sp}(6,2)|= 2903040$ maximum cliques of size $4$ in the dual graph $\mathcal{G}_{2^3 \times 3}^*$ corresponding to the same number of incomplete sets of mutually unbiased bases. The maximum cliques of the Pauli graph intersect each other at $0$, $1$, $2$, $3$, $5$ or $11$ points and the $5$-tuples form a graph of spectrum $\{56^1,14^{15},2^{35},-4^{84}\}^4$, related to three copies of the polar space $W_5(2)$. The automorphism group of this geometry is found to be $\mbox{aut}(\mathcal{G}_{2^3 \times 3}^{(5)})=\mbox{Sp}(6,2)^4.S_4$, a straightforward generalization of what occurs for the $2$-qubit/qutrit system.

\section{Conclusion}

It has been shown for the first time that number theoretical functions $\sigma(q)$ and $\psi(q)$ enter into the structure of commutation relations of Pauli graphs and geometries. For single $q$-dits (in section 2), $\sigma(q)$ and $\psi(q)$ refer to the number of maximal commuting sets and the cardinality of the projective line $\mathbb{P}_1(\mathbb{Z}_q)$, respectively. For multiple qudits, with dimension $p^n$, $p$ a prime number, (in section 3) the parameter $q=p^{2n-1}$ enters in the function $\sigma(q)$ to count the size of the symplectic polar space $W_{2n-1}(p)$ (that carries the multiple qudit system), and enters in the function $\psi(q)$ to count the size of the basic constituent: the punctured polar space $W'_{2n-1}(p)$. For multiple qudit mixtures, spaces $W_{2n-1}(p)$ and  $W_{2n-1}'(p)$ are also found to arise as constituents of the commutation structure.

The structural role of symplectic groups $Sp(2n,p)$ has been found, as expected. Other important symmetry groups are $G_{48}=\mathbb{Z}_2\times S_4$, $G_{144}=A_4 \times D_6$ and $W(E_6)$. The group $G_{48}$ is first of all the automorphism group of the single qudit Pauli group $\mathcal{P}_1$ and is important in understanding the CPT symmetry \cite{PlanatCPT}. In this paper, it arises as the symmetry group of the quartit, of the punctured generalized quadrangle $GQ(2,2)'$ (see  \ref{dGQ22})) and as a normal subgroup of many systems of qudits (as shown in Table 1). The torus group $G_{144}$ occurs in the symmetries of the $6$-dit, $12$-dit, $18$-dit and $24$-dit systems. The Weyl group $W(E_6)$ happens to be central in the symmetries of three-qubit and multiple qutrit systems. The understanding of symmetries in the Hilbert space is important for the applications in quantum information processing. 

%In a next paper, we intend to display the topological properties of the Pauli graphs, viewing them as simplicial complexes. 

\section*{Acknowledgements} The author thanks Fabio Anselmi for his feedback on this topic and for his critical reading of the manuscript.

\end{document}